\documentclass[aps,prl,twocolumn,superscriptaddress,floatfix]{revtex4-1}
\usepackage{graphicx,amsmath,amssymb,amsxtra}
\usepackage{color}
\def\bd{\begin{displaymath}}
\def\be{\begin{equation}}
\def\ed{\end{displaymath}}
\def\ee{\end{equation}}
\def\bsub{\begin{subequations}}
\def\esub{\end{subequations}}

\usepackage{tikz}
\usetikzlibrary{shapes}

\newcommand{\Eq}[1]{Eq.~(\ref{#1})}
\newcommand{\Fig}[1]{Fig.~\ref{#1}}

\makeatletter
\newcommand*\dashline{\rotatebox[origin=c]{90}{$\dabar@\dabar@\dabar@$}}
\makeatother

\def\Eq#1{Eq. {\eqref{#1}}}

\def\be{\begin{equation}}
\def\ee{\end{equation}}

\def\bea{\begin{eqnarray}}
\def\eea{\end{eqnarray}}

\usepackage{subfigure}
\usepackage{times}
\usepackage{graphicx}
\usepackage{xcolor,dsfont}

\graphicspath{{figures}}

\begin{document}
\title{Exact Solution and Correlations of a Quantum Dimer Model on the Checkerboard Lattice}

\author{Julia Wildeboer}
\email{Julia.Wildeboer@asu.edu}
\affiliation{Max-Planck-Institute for the Physics of Complex Systems, 01187 Dresden, Germany}
\affiliation{Department of Physics, Arizona State University, Tempe, Arizona 85287-1504, USA}

\author{Zohar Nussinov}
\email{zohar@wustl.edu}
\affiliation{Department of Physics, Washington University, St. Louis, Missouri 63130, USA}

\author{Alexander Seidel}
\email{seidel@physics.wustl.edu}
\affiliation{Department of Physics, Washington University, St. Louis, Missouri 63130, USA}

\begin{abstract}
We present analytic results for a special dimer model on the {\em non-bipartite} and {\em non-planar}  
checkerboard lattice that does not allow for parallel dimers surrounding diagonal links.  
We {\em exactly} calculate the number  of closed packed dimer coverings on finite 
checkerboard lattices under periodic boundary conditions, and determine all dimer-dimer correlations. 
The latter are found to vanish beyond a  certain distance. We find that this solvable model, despite being non-planar, 
is in close kinship with well-known paradigm-setting planar counterparts that allow exact mappings to 
$\mathbb{Z}_2$ lattice gauge theory.  
\end{abstract}

\maketitle

{\em Introduction.} -- 
A major challenge in condensed matter theory lies in the endeavor of  
finding relatively simple toy models that are tractable and at the same time 
capture relevant universal physics. 
One long standing example for this are magnetic systems whose collective behavior  
can often be well-described by relying on simple Ising- or Heisenberg-type models. 
Even the latter are, however, beyond analytic treatment in all but the simplest possible settings. 
The need for solvable models is particularly acute when investigating physics outside established paradigms. 

Historically, models with dimer degrees of freedom (and their associated constraints) 
have played a key role in constructing solvable models in statical physics and, somewhat more recently, 
quantum magnetism. 
The enormous impact dimer models have had on various areas of theoretical physics 
can be traced back to Kasteleyn's observation that a large class of classical dimer models 
is solvable by Pfaffian methods on planar lattice graphs \cite{kas}, 
followed by developments by Fisher \cite{Fisher} and 
Fisher and Stephenson \cite{FS}, revealing deep connections with Ising models. 
This method has since been adopted to shine light on the phase diagram of quantum magnets. 
Kivelson, Rokhsar, and Sethna \cite{RKS,RK} have introduced the idea that quantum dimer models 
(QDMs) are effective descriptions of highly frustrated quantum magnets, 
and can be tuned such that their ground-state correlations correspond to 
those of a classical dimer model \cite{RK}. 
This, in particular, is linked to scenarios of unconventional magnetism 
conceived during the advent of high-temperature superconductors \cite{RKS,anderson1987resonating,fazekas1974ground,Fradkin1990}. 
In a seminal work, Moessner and Sondhi \cite{ms} demonstrated that Anderson's idea of a short-ranged 
resonating valence bond spin-liquid phase can be realized, under the assumption of validity 
of the spin-to-quantum-dimer mapping, on the triangular lattice. 
This assumption has been corroborated via a multitude of different approaches, 
including systematic expansion in an overlap parameter \cite{expansion}, which rests 
on the linear independence \cite{CCK,wildeboer11} 
of spin singlet (valence bond) states, 
and the construction of $SU(2)$-invariant spin-$1/2$ models  
that realize the same ground states \cite{seidel09,normand14} 
and their physics \cite{wildeboer12,wildeboer15,wildeboer17}. 
Moreover, deep connections between certain QDMs and Kitaev's toric code for topological quantum computing \cite{kitaev} 
have long been appreciated, and the relation to the underlying Ising ($\mathbb{Z}_2$) gauge theory can be made 
exact in the kagome lattice model discussed by Misguich et. al \cite{misguich}. 
Other gauge theories describing QDMs on different lattices have been introduced and 
studied \cite{Fradkin1990,Moessner2001,Nogueira2009}. 
QDM type physics also appears in various orbital and spin-orbital systems \cite{Vernay06} 
and Josephson junction arrays \cite{Albuquerque07}. 

Up until now, the construction of dimer models has thus proven a profound and versatile tool, 
whose utility in the applications discussed above was, however, largely limited to planar lattice graphs. 
In this work, we show that no such limitation fundamentally exists. 
That is, we construct a dimer model on the checkerboard lattice, 
which is non-planar due to crossing links, and has all of the benefits discussed above. 
In this work, we demonstrate the applicability of Pfaffian methods to this model. 
The model further allows exact mapping to Ising gauge theory, along with 
existences of pertinent local lattice symmetries, 
on which we will elaborate elsewhere. 

\begin{figure}
\includegraphics[width=0.45\columnwidth]{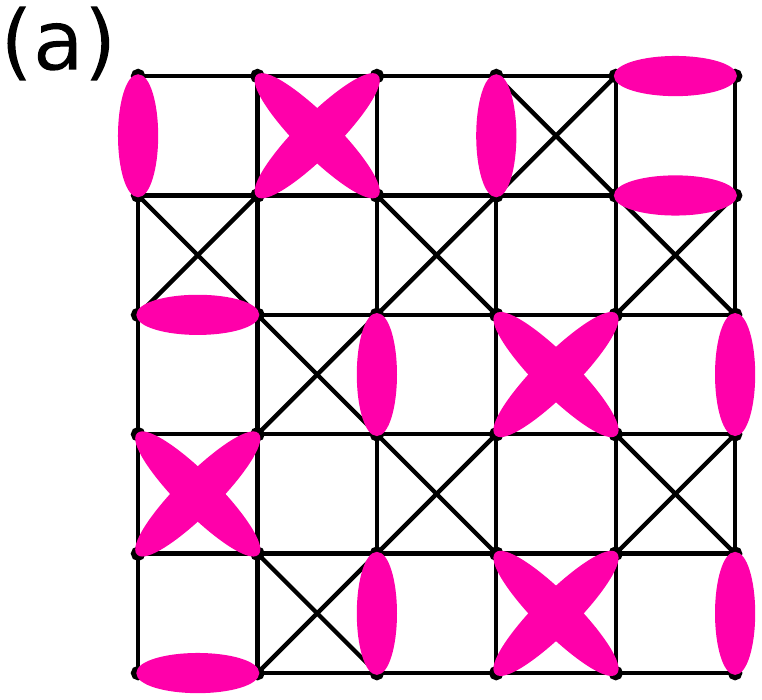} \quad \quad
\includegraphics[width=0.45\columnwidth]{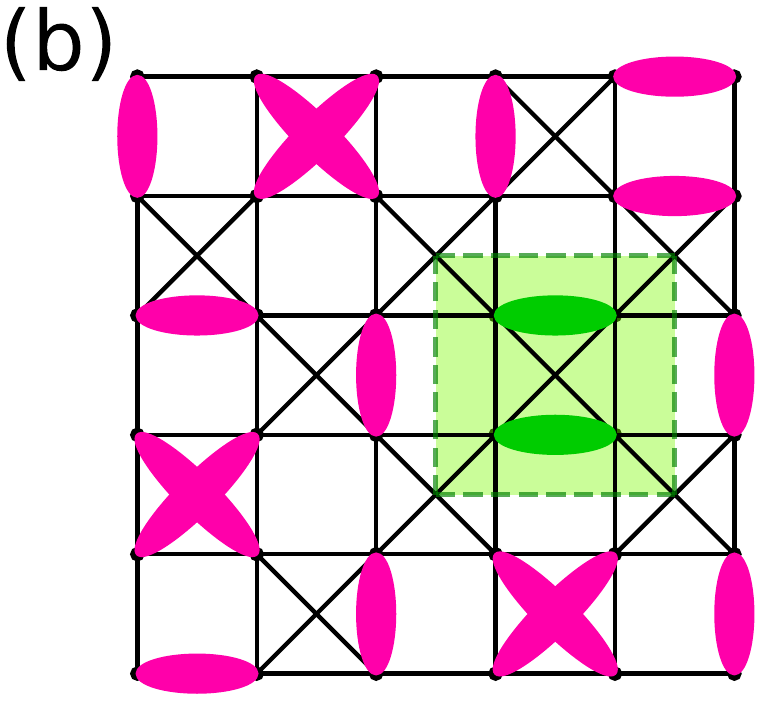}
\caption{The checkerboard lattice with some dimer coverings. 
        (a) exemplifies a dimer covering that satisfies all constraints of our model. 
        (b) A dimer covering that features a pair of parallel dimers (shaded) on the 
            same crossed plaquette and is thus not allowed. 
}
\label{figure2}
\end{figure}

{\em A non-planar dimer model.} -- 
We will now introduce a dimer model 
acting on the space of {\em restricted} dimer coverings on the checkerboard lattice (Fig. ~\ref{figure2}).  
The model can be interpreted as a classical dimer model at infinite temperature, 
but at the same time, the correlations to be discussed have equal relevance to the 
ground state of a suitable QDM. 
The configuration space of this model is that of all possible 
dimer coverings subject to an additional constraint. 
A ``dimer covering'' refers to a placing of dimers on some of the links of a lattice, 
such that each vertex  belongs to exactly one such dimer. 
Here, crossed dimers on crossed-linked plaquettes are explicitly allowed. 
However, we impose the restriction that cross-linked plaquettes may not admit a pair of 
parallel dimers (Fig.~\ref{figure2} and caption). 
Effectively, we thus introduce an interaction ascribing a very large energy to pairs 
of dimers occupying the vertical or horizontal links on a cross-linked plaquette, 
whereas there is no such penalty for single {\em or} double occupation of crossed links. 
In the following, we will set this energy penalty to infinity first, 
and then consider the infinite temperature partition sum of the resulting 
constrained dimer model. 

It is worth pointing out that constrained dimer models on non-planar lattices 
have appeared in the literature before \cite{yao2012exact}, though typically, 
the constraint disallows crossed dimers. 
Even then, the non-planarity of the model is evident by the fact that typical overlap 
graphs \cite{FradkinBook} between two different dimer coverings will have many crossings. 
We are not aware, however, of any such model that allows exact calculation of 
partition/correlation functions, which is what we will now turn to for the model at hand. 
To appreciate why that is possible, we first make contact with a well-known theorem by 
Kasteleyn \cite{kas}, which, however, applies only to planar lattices. 

\begin{figure}[t]
\includegraphics[width=0.99\columnwidth]{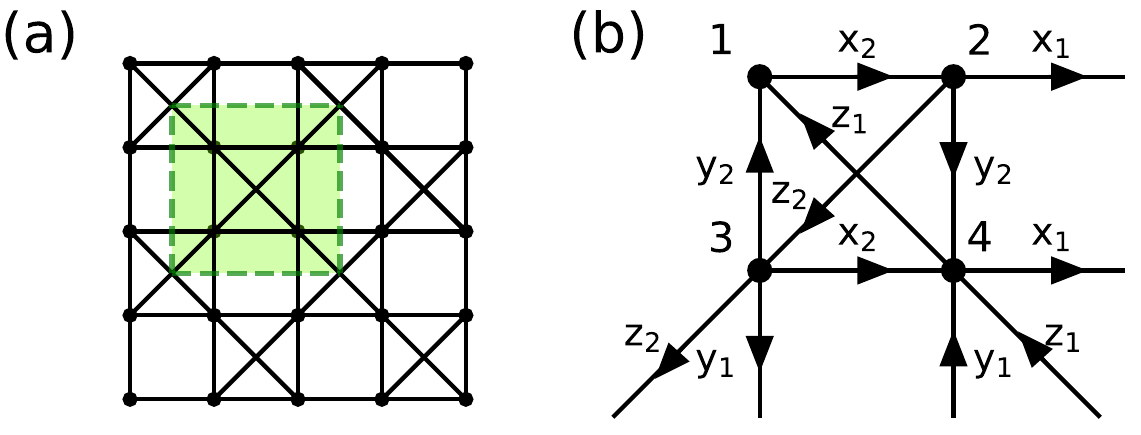}
\caption{
        (a) The checkerboard lattice is shown, the unit cell is depicted 
        inside the dashed lines. (b) shows the four-site unit cell of the 
        checkerboard lattice. It contains twelve links each equipped with a respective 
        weight of $x_{1}, x_{2}, y_{1}, y_{2}, z_{1}, z_{2}$. 
        The arrows on the links indicate the Kasteleyn orientation. 
}
\label{figure1_666}
\end{figure}

According to this theorem, for any planar graph, an antisymmetric matrix $A$ can be found 
such that the infinite temperature dimer partition function   
is given by the Pfaffian ($\rm Pf$) of the matrix $A$. 
This matrix encodes a link orientation of the graph, which has 
certain additional properties making it a ``Kasteleyn orientation''. 
Here, the indices of $A$ are site indices of the lattice, and 
$A_{ij}\neq 0$ only if $i$, $j$ are connected by a link. 
The orientation is defined by arrows placed along the links (Fig. ~\ref{figure1_666}(b)), 
and $A_{ij}=1$ if the arrow points from $i$ to $j$, $A_{ij}=-1$ if it points from $j$ to $i$.  
The defining property of a Kasteleyn orientation is to place arrows on the links of the planar  
lattice so that each plaquette is ``clockwise odd'', e.g. the number of clockwise arrows around  
any elementary plaquette (face) is odd. 
Ref.[\onlinecite{wildeboer2017entanglement}] introduced the notion of a ``pre-Kasteleyn'' 
orientation. This notion is meaningful even for non-planar two-dimensional lattices, 
i.e., lattices equipped with crossing links. 
An orientation for such links was defined to be ``pre-Kasteleyn'' if for any closed, 
non-self-intersecting, contractible loop along links, the number of clockwise oriented links 
is even if the number of sites enclosed by the loop is odd, and vice versa. 
The term ``contractible'' is necessary only in the presence of non-trivial boundary conditions, 
in particular toroidal, periodic ones. 
The main difference between Kasteleyn and pre-Kasteleyn orientations is that for the latter, 
we do not need well-defined notions of an ``elementary plaquette'', or face, of the lattice.  
We only need the lattice to be meaningfully embedded within a two-dimensional planar or toroidal surface, 
so that ``enclosed'' is well-defined.  
The two notions agree, however, for planar lattice graphs \cite{wildeboer2017entanglement}. 
The arrows shown in Fig. ~\ref{figure1_666}(b) do endow the checkerboard lattice with a pre-Kasteleyn orientation. 
This follows from the fact that the checkerboard is obtained form the frustrated square lattice (with has all square cross-linked) 
and its pre-Kasteleyn 
orientation discussed in 
Ref.[\onlinecite{wildeboer2017entanglement}] via removal of links. 

For an ordinary Kasteleyn orientation, and assuming open boundary conditions, 
one has that ${\rm Pf} (A)$ equals (up to a sign) the number of all dimer coverings, i.e., the infinite temperature 
dimer partition function. Similarly, for our pre-Kasteleyn orientation, ${\rm Pf} (A)$ can be written as as sum, over dimer 
coverings, of terms $\pm 1$. Here the positive (negative) sign corresponds to dimer coverings with an even 
(odd) number of crossed dimers. 
We now define ``physical'' dimer coverings as those that do not have parallel dimers on any cross-linked 
plaquette. On the checkerboard lattice, any un-physical dimer covering is uniquely associated to a physical 
one, by replacing all un-physical parallel pairs with crossed dimers on the same plaquette. 
If we now consider any pair of crossed dimers together with the two un-physical configurations 
associated to it, 
we find that these three local configurations contribute $-1+1+1=1$ to ${\rm Pf} (A)$. 
One then easily realizes that, for checkerboard lattices of any size and shape (open boundary conditions), 
${\rm Pf} (A)$ gives the number of physical dimer configurations. 
Such counting problems certainly have great tradition in the field \cite{kas,Fisher}. 
One has the intuition that whenever this counting is possible, then, at the very least, correlations can also be calculated.  
This will indeed turn out to be the case. 
 
We proceed by reviewing nuts and bolts of Kasteleyn's formalism. 
Assuming, now, periodic boundary conditions, the partition sum for all dimer coverings is given 
by\footnote{This statement carried over from the Kasteleyn orientation of the 
square lattice \cite{kas} to our checkerboard pre-Kasteleyn orientation.} 
\begin{eqnarray}\label{pbc_pfaf}
Z &=& \frac{1}{2}\left(-{\rm Pf}(A^{00}) + {\rm Pf}(A^{\frac{1}{2}0}) + {\rm Pf}(A^{0\frac{1}{2}}) 
+ {\rm Pf}(A^{\frac{1}{2}\frac{1}{2}})\right)\,,\nonumber\\
&&
\end{eqnarray}
where $A^{00}$ encodes the (pre)-Kasteleyn orientation as before, in the presence of 
periodic boundary conditions. The other three matrices 
are the same, except for the presence of vertical and/or horizontal boundary ``twists''. 
Here, a twist introduces a flip of orientation along all links crossed by a closed path that 
traverses the lattice horizontally (vertically) as indicated by a $\frac 12$ in the first (second) index. 
The Pfaffians ${\rm Pf}(A^{\alpha\beta})$ then satisfy 
\begin{eqnarray}\label{pbc_det}
{\rm Pf}(A^{\alpha\beta}) = \pm \sqrt{{\rm det}(A^{\alpha\beta})}\;,
\end{eqnarray}
where we will  fix the signs below.  
The matrices $A^{\alpha\beta}$ can be block-diagonalized by a Fourier transformation, giving 
\begin{eqnarray}\label{ev_det}
	{\rm det}(A^{\alpha\beta}) = \prod_{n=0}^{N-1}\prod_{m=0}^{M-1} {\rm det}A(\theta_{n,\alpha},\phi_{m,\beta}),
\end{eqnarray}
where the matrix-blocks $A(\theta_{n,\alpha},\phi_{m,\beta})$ are given by \cite{breiman1964stopping} 
\begin{eqnarray}\label{det_lam}
A(\theta_{n,\alpha},\phi_{m,\beta}) = \sum_{M_1,N_1}a_{M_1,N_1}e^{i(N_{1}\theta_{n,\alpha}+M_{1}\phi_{m,\beta})}\,.
\end{eqnarray}
Here, the phases $\theta_{n,\alpha}$ and $\phi_{m,\beta}$ are specified via 
\begin{eqnarray}
\theta_{n,\alpha} = \frac{2\pi(n + \alpha)}{N}\;\;\;\;\text{and}\;\;\;\;\theta_{m,\beta} = \frac{2\pi(m + \beta)}{M}
\end{eqnarray}
with $n = 0,\ldots,N-1$ and $m = 0,\dots,M-1$. $M$ $(N)$ is the number of unit cells as depicted in Fig.~\ref{figure1_666}(b) 
in vertical (horizontal) direction. 
The $a_{M_1,N_1}$ encode the Kasteleyn orientation as follows: 
Let $j=(M_1,N_1,\nu)$ be a multi-index specifying the lattice site 
in the unit cell given by indices $M_1$, $N_1$ and corresponding 
to a unit-cell basis-index $\nu$, and let similarly $i=(0,0,\mu)$ 
specify a lattice site in the $(0,0)$-unit cell, then 
$[a_{M_1,N_1}]_{\mu,\nu}=A_{ij}$. Formally, 
$M_1$ and $N_1$ also run over  $M$ and $N$ distinct values, 
respectively, but only values $M_1,N_1=-1,0,1$ will lead to nonzero 
$a_{M_1,N_1}$, referring to a unit cell $(0,0)$ and its neighbors. 

The pre-Kasteleyn orientation of \Fig{figure1_666}(b) does, by itself, not enlarge the two-site unit cell of the 
checkerboard lattice. However, we may be interested in a more general problem by endowing links 
with certain positive weights $x_k$, $y_k$, $z_k$, $k=1,2$, as shown in the figure. The weights multiply 
the corresponding matrix elements of $A_{ij}$. 
For 
$x_1^2+y_1^2\geq z_1z_2$, $x_2^2+y_2^2\geq z_1z_2$ 
the identification of unphysical parallel dimer pairs with crossed pairs 
can still be interpreted as a positive partition function. The resulting unit cell then has four sites. Though in the end, we 
mostly will be interested in the case with all weights equal to $1$, it has certain advantages to think 
of the larger unit cell whose sites comprise one crossed plaquette, garnished with the weights shown in \Fig{figure1_666}(b). 
In the following, we will refer to this unit cell, which contains twelve links. 

The nine non-zero matrices 
 $a_{M_{1},N_{1}}$ 
may now be read off from \Fig{figure1_666}(b). One has 
\begin{eqnarray}
 a_{0,1} & = & \begin{pmatrix}
 0   &  0 & 0 & 0\cr
 x_1 &  0 & 0 & 0\cr
 0   &  0 & 0 & 0\cr
 0   &  0 & x_1 & 0\cr
\end{pmatrix},\quad
a_{1,0} = \begin{pmatrix}
 0   &  \;\;\;0 & 0 & 0\cr
 0   &  \;\;\;0 & 0 & 0\cr
 y_1 &  \;\;\;0 & 0 & 0\cr
 0   &  -y_1 & 0 & 0\cr
\end{pmatrix}, \nonumber\\
a_{1,1} & = & \begin{pmatrix}
 \;\;\;0   &  0 & 0 & 0\cr
 \;\;\;0 &  0 & 0 & 0\cr
 \;\;\;0   &  0 & 0 & 0\cr
 -z_1   &  0 & 0 & 0\cr
\end{pmatrix} ,\quad
a_{1,-1}   =  \begin{pmatrix}
 0   &  0 & 0 & 0\cr
 0   &  0 & 0 & 0\cr
 0   &  z_2 & 0 & 0\cr
 0   &  0 & 0 & 0\cr 
\end{pmatrix}, \\
a_{0,0} & = & \begin{pmatrix}
\;\;\;0 &  \;\;\;x_2 & -y_2 & -z_1\cr
   -x_2 &  \;\;\;  0 & \;\;\;z_2 & \;\;\;y_2\cr
\;\;\;y_2 & -z_2 &\;\;\; 0 & \;\;\;x_2\cr
\;\;\;z_1 & -y_2 & -x_2 & \;\;\;0\cr
\end{pmatrix} \;.\nonumber
\label{equ:amatrices}   
\end{eqnarray}
Furthermore,
\begin{eqnarray}	
&&a_{-1,0} = - a_{1,0}^{T} ,\quad a_{0,-1}   =   -
a_{0,1}^{T}, \nonumber\\
&&a_{-1,-1}  = - a_{1,1}^{T},\quad a_{-1,1}  = - a_{1,-1}^{T}. 
\end{eqnarray}
The matrices $A(\theta_{n,\alpha},\phi_{m,\beta})$ can be written as 
\begin{widetext}
\begin{eqnarray}\label{kas_matrix}
A(\theta_{n,\alpha},\phi_{m,\beta}) &=& \sum_{M_1,N_1}a_{M_1,N_1}e^{i(N_{1}\theta_{n\alpha}+M_{1}\phi_{m\beta})}\nonumber \\
&=& a_{0,0} + a_{0,1}e^{i\theta_{n}} - a_{0,1}^{T}e^{-i\theta_{n}} + a_{1,0}e^{i\phi_{m}} - a_{1,0}^{T}e^{-i\phi_{m}} 
            + a_{1,1}e^{i(\theta_{n}+\phi_{m})} - a_{1,1}^{T}e^{-i(\theta_{n}+\phi_{m})} \nonumber \\
& & + a_{-1,1}e^{-i(\theta_{n}-\phi_{m})} - a_{-1,1}^{T}e^{i(\theta_{n}-\phi_{m})} \nonumber \\
&=&    
\begin{pmatrix}
        0                                 &  x_2 - x_1 e^{-i\theta_n}           & -y_2 - y_1 e^{-i\phi_m}          & -z_1 + z_1 e^{-i(\theta_n+\phi_m)}\cr
        -x_2 + x_1 e^{i\theta_n}          &  0                                  & z_2 - z_2 e^{i(\theta_n-\phi_m)} & y_2 + y_1 e^{-i\phi_m}\cr
        y_2 + y_1 e^{i\phi_m}             &  -z_2 + z_2 e^{-i(\theta_n-\phi_m)} & 0                                & x_2 - x_1 e^{-i\theta_n}\cr
        z_1 - z_1 e^{i(\theta_n+\phi_m)}  &  -y_2 - y_1 e^{i\phi_m}             & -x_2 + x_1 e^{i\theta_n}         & 0\cr
\end{pmatrix}.
\end{eqnarray}\;
\end{widetext}
Any single block \eqref{kas_matrix} is, in general, not skew-symetric and so by itself does not represent a well-defined contribution 
to the Pfaffian. However, such a block comes with a conjugate partner, and a change of basis within the two blocks 
(corresponding to a real, sine-cosine Fourier transform of the original matrix) restores skew symmetry, and shows that such 
partners contribute a positive factor to the Pfaffian 
\footnote{The determinant overall matrix transformation is taken into account in all pertinent arguments.}. 
For even $M,N$, except in $A^{00}$, all blocks \eqref{kas_matrix} come with conjugate partners, 
so the Pfaffians of the remaining three $A^{\alpha\beta}$ 
lead to a positive contribution. 
For $A^{00}$, the sign of the Pfaffian may be worked out from the four special, 
already skew symmetric blocks with $\theta, \phi \in \{0,\pi\}$. 
One then finds that it is given by the sign of 
\begin{eqnarray}\label{helper1}
&&\left[(x_1+x_2)^2+(y_1+y_2)^2-4z_1z_2\right] \times \nonumber\\
&&\left[(x_1-x_2 )^2 +(y_1-y_2)^2-4z_1z_2\right]\,,
\end{eqnarray}
where the first factor is non-negative under our earlier assumption, $x_1^2+y_1^2\geq z_1z_2$. 
In particular, if we now specialize to $x_1 = x_2 = x, y_1 = y_2 = y$ and $z_1 = z_2 = z$ for simplicity, 
the sign of ${\rm Pf}A^{00}$ is negative, \Eq{pbc_pfaf} becomes 
\begin{eqnarray}\label{pbc_psum}
	Z = \frac{1}{2} \sum_{\alpha \beta} \sqrt{{\rm det}(A^{\alpha \beta})}\;\;\;\text{for}\;\; \alpha,\beta = 0,\frac{1}{2}\;. 
\end{eqnarray}
From \eqref{kas_matrix}, 
\begin{eqnarray}\label{detA}
{\rm det}A(\theta_{n,\alpha},\phi_{m,\beta}) = 4[x^2 + y^2 &+& (z^2 - x^2){\rm cos \;}\theta \nonumber\\ 
					&+& (y^2 - z^2){\rm cos \;}\phi]^{2}\;.
\end{eqnarray}
In the thermodynamic limit, the distinction between ``twists'' $(\alpha,\beta)$ becomes irrelevant, 
and we may evaluate the per-dimer free energy as 
\begin{widetext}
\begin{eqnarray}\label{f_int}
&f(x, y, z)
=  \frac{1}{2(2\pi)^{2}} \int_{0}^{2\pi} \int_{0}^{2\pi} 
        {\rm ln \;Pf}[A(\theta, \phi)] \,d\theta \,d\phi \nonumber 
        \\ &=\frac{1}{2\pi} \int_0^{2\pi} \log [x^2 + y^2 + (z^2-x^2) \cos\theta + \sqrt{  (x^2 + y^2 + (z^2-x^2) \cos \theta)^2 -(z^2-y^2)^2 }] \, d \theta. 
\end{eqnarray}
\end{widetext}
It is interesting to evaluate this last expression in the limit $z^2\rightarrow x^2+y^2$, 
when the effective weight of crossed dimers becomes zero. 
In this limit we have a highly constrained model, where diagonal dimers are still possible,   
but no pair of dimers, parallel or crossed, may occupy a cross-linked plaquette. 
One may infer from the last equation that this is a critical point, where 
\begin{equation}
f = \frac{x^2+y^2-z^2}{2\pi(y^2-z^2)^2}\log(x^2+y^2-z^2)+\dotsc,
\end{equation}
and where the ellipses represent less singular terms. 
It is interesting to note that field-theoretic mappings indicate 
an abundance of (first-order) transitions in the phase diagram of the 
frustrated square lattice (with one diagonal in each square) \cite{Trousselet}. 
The present model gives analytic access to similar transitions in a microscopic setting. 

Specializing from now on to $x=y=z$, and going back to a finite lattice with PBCs, Eqs. \eqref{kas_matrix}-\eqref{detA} 
directly give 
\begin{eqnarray}
	Z= 2 \cdot (4x^{2})^{MN} \overset{x = 1}{=}  2 \cdot 4^{MN}\;,	
\end{eqnarray}	
where the $x=1$ result is the number of dimer coverings of the lattice of $MN$ (four-site) unit cells \footnote{
One may similarly show that all four ``topological sectors'' in the precence of PBCs contribute equally to this number.}. 
Interestingly, the latter is formally the same expression as for the kagome. We caution, however, 
that the number of sites per unit cell is different for the checkerboard and kagome, so the counting in terms 
of lattice sites is different. 

{\em Correlations.} -- 
Though the present case represents a highly 
non-standard application of Pfaffian methods,  
experience nonetheless suggests that if $Z$ 
is computable, then so are correlation functions. 
We will now show that this is indeed the case. We are interested in the 
correlation of the dimer operator $n_{ij}$, where $i$ and $j$ denote neighboring lattice sites, 
and $n_{ij} = 1(0)$ if the link $ij$ is occupied (empty). 
Since products of these operators are projection operators, their expectation values 
can be written as $Z^{\prime}/Z$, where $Z$ is the original partition function, 
and $Z^{\prime}$ the partition function restricted to the subspace onto which the 
operator in question projects. 
In practice, $Z^{\prime}$ is the partition function of the same lattice with certain links removed. 
E.g., if ${ij}$ is a horizontal or vertical link, one may easily see that dimerizations 
that have this link occupied, subject to our no-double-occupancy rule, 
are in one-to-one correspondence with dimerizations of the same lattice that have all 
{\em other} links of the cross-linked square containing $ij$ removed, as well as all other links 
attaching to either $i$ or $j$. For a diagonal link $ij$, the same prescription 
effectively leads to counting all dimerizations having this link occupied {\it but not crossed}. 
Let's call the associated partition function $Z^{\prime\prime}$.  
The partition function $Z^{\prime\prime\prime}$ of configurations 
where $ij$ is occupied {\it and} crossed is similarly related to the partition function of the lattice 
with all links on or attached to the cross-linked plaquette removes, {\em except} for the cross\footnote{
A factor of $(-1)$ must be accounted for due to the presence of the cross}. 
 Thus $Z^{\prime}=Z^{\prime\prime}+Z^{\prime\prime\prime}$. 
Products of $n_{ij}$ operators are dealt with accordingly. 
Based on these observations, the calculation of correlation functions for the present problem 
differs only slightly from the standard case of unrestricted dimer coverings of a planar  
lattice graph. The difference is only in working out the links to be removed for a given  
numerator $Z^{\prime}$. We may denote by $\Delta$ the matrix obtained from $A$ by keeping only 
those matrix elements corresponding to removed links, setting the others equal to zero. 
It is then standard to express $Z^{\prime}/Z$ (or $Z^{\prime\prime}/Z$, $Z^{\prime\prime\prime}/Z$) 
in terms of $\Delta$ and the Green's function matrix, $G = A^{-1}$. 
We review this technique in the Supplemental Material. 

We are now interested in connected correlation functions 
\begin{eqnarray}\label{corrs}
C[ij,kl] = \langle n_{ij}n_{kl}\rangle 
- \langle n_{ij}\rangle \langle n_{kl}\rangle\;.
\end{eqnarray}
Classical and quantum dimer models throughout the literature exhibit a great variety of behaviors, 
including power-law \cite{leung,dimersquare,huse} 
and (super)-exponentially decaying \cite{ms,misguich} 
correlations, mirrored by a class of closely related spin-degree wave functions \cite{misguich,fujimoto,sutherland}. 
In the present case, we find the correlations \eqref{corrs} to be ultra-short ranged, i.e., 
non-vanishing only up to a certain finite distance. The finitely many non-zero values 
of the correlator are listed in Tables ~\ref{table1}-\ref{table3} of the Supplemental Material. 
This property is familiar from a few select dimer models, notably that on the kagome lattice \cite{misguich}. 
It hints at a deeper solvable structure of the present model, which we now outline. 

{\em Discussion and conclusion.} -- 
Planar dimer models exhibit a plethora of phases, including broken symmetry and 
$\mathbb{Z}_{2}$-topological phases. 
The correlators addressed in the preceding section do not indicate any broken symmetry. 
On the other hand, with periodic boundary conditions, dimer coverings can be sub-divided into 
four topological sectors, as familiar from planar dimer models \cite{FradkinBook}. 
The latter transform non-trivially under symmetries of the lattice. Related to this, any absence of symmetry 
breaking in dimer models has long been associated to topological order \cite{ReadChakraborty,Kivelsonpaper}. 
Indeed, these arguments may be sharpened when considering quantum dimer models of the RK-type, whose 
ground state correlators agree with those of the classical model considered here. In particular, 
this then allows one to study question of universality through entaglement properties 
of the ground state, and properties of the excitations, such as braiding statistics. 
Such a program can be carried out in full detail for the present model. 
Here we summarize key features, while details will be given elsewhere \cite{wns_checker}. 
 
It is worth noting that a small subset of QDMs are fully solvable -- all eigenstates are known, not just 
the RK ground state. This is in particular true for the kagome QDM \cite{misguich}, which can be written 
as the sum of commuting local operators, permitting an exact mapping to Ising gauge theory, 
and for which the vanishing of all correlations between 
local operators at sufficient distance can be 
demonstrated exactly \cite{misguich,  Oshikawa, wang2007exact,wu2008dimers}. 
Despite its non-planarity, we have noted a number of parallels between the present model and the kagome case.  
This is no coincidence. The key uniting feature between these models turns out to be the existence of an 
arrow representation for permissible dimer coverings, which has been appreciated for the kagome for some time \cite{Elser}. 
This translates the construction principle for the kagome QDM to the present case, with all the benefits mentioned. 
Moreover, the calculation of ground state entanglement entropy is possible {\it exactly}, 
exposing a topological part of $\ln 2$, proving the topological nature of the ground state.   
Finally, quasiparticle statistics are accessible through modular properties of so-called minimally entangled 
states (MES) \cite{vishvanath}, which are again exactly computable for the QDM associated to the present model.   

The purpose of this work is to illustrate that a wealth of beautiful models realizing topological orders lies 
hidden in non-planar dimer physics. Such models can be made accessible through the notion of a pre-Kasteleyn 
orientation. We have discussed a checkerboard model that is fully solvable and whose quantum version describes 
a topological liquid. We are hopeful that this approach will stimulate many fruitful developments. 
 
\bibliography{bibfile} 

\begin{widetext}

\begin{center}
\bf{\large Supplemental Material for ``Exact Solution and Correlations of a Quantum Dimer Model on the Checkerboard Lattice''}
\end{center}
\section{Dimer expectation values and correlations}\label{AppendixA}

We will obtain the short-distance connected correlations between all twelve links contained in the 
four-site unit cell of the quantum dimer model on the checkerboard lattice. 
We copy \Eq{corrs} from the main text, 
\begin{eqnarray}\label{corrs2a}
C[ij,kl] = \langle n_{ij}n_{kl}\rangle 
- \langle n_{ij}\rangle \langle n_{kl}\rangle\;.
\end{eqnarray}
We will be interested in the thermodynamic limit, and so specialize the discussion to the case of 
open boundary conditions. Periodic boundary conditions can be treated similarly. 
By translational invariance, the correlations \eqref{corrs2a} only depend on relative distance. 
We will now show how to obtain the expectation values of operators of type $n_{ij}n_{kl}$ and $n_{ij}$. 
As explained in the main text, these expectation values 
can be written as $Z^{\prime}/Z$, where $Z$ is the original partition function, 
and $Z^{\prime}$ the partition function restricted to the subspace where the 
operator in question equals $1$. 
In turn, $Z'$ can be re-interpreted in terms of one or more partition functions 
corresponding to lattice graphs similar to the original one, but with several links removed. 
E.g., for the case  of the expectation value of a single link $ij$ that is horizontal/vertical, $Z'$ equals 
the partition function of the a lattice with all links on the same cross-linked plaquette as $ij$ removed, 
other than $ij$ itself, and also all links that touch either $i$ or $j$. 
Similarly, for a diagonal link, $Z'=Z''+Z'''$, where $Z''$ and $Z'''$ {\em each} are defined by a removal of a 
different set of links, as explained in the main text. We focus on the case where $Z'$ can be defined as the 
partition function corresponding to the removal of one particular set of links, 
as is the case for horizontal/vertical dimers,  with other cases reducing to sums of such partition functions. 
This is in particular also true for the general case of products $n_{ij}n_{kl}$ of dimer operators. 
This then  
translates into the equation 
\begin{eqnarray}
	Z^{\prime} = \textrm{Pf}A^{\prime} = \textrm{Pf}[A+\Delta]\;,
\end{eqnarray}
where, if $p$ links $(m,n), \ldots, (k,l)$ are removed,  
then $\Delta$ is a matrix of the same dimension as $A$  
with only $2p$ non-zero elements, $\Delta_{m,n}=-A_{m,n}$,  $\Delta_{n,m}=-A_{n,m}$, $\ldots$, 
$\Delta_{k,l}=-A_{k,l}$, $\Delta_{l,k}=-A_{l,k}$. 
Consequently, we have 
\begin{eqnarray}
	\langle n_{ij}\rangle = Z^{\prime}/Z = \frac{\textrm{Pf}A^{\prime}}{\textrm{Pf}A} = \frac{\textrm{Pf}[A+\Delta]}{\textrm{Pf}A}
\end{eqnarray}
and 
\begin{eqnarray}
\langle n_{ij}\rangle^{2} = \frac{\textrm{det}A^{\prime}}{\textrm{det}A} = \frac{\textrm{det}[A(I+G\Delta)]}{\textrm{det}A} = \textrm{det}(I+G\Delta)\;.
\end{eqnarray}
Here $I$ is the identity matrix and $G \equiv A^{-1}$ is the Green's function matrix. 
The procedure outlined so far is applicable to any finite lattice with arbitrary dimer weights. 
We now turn to the special case of a large lattice and determine the exact expression 
for the elements of $G_{ij}$ for uniform weights 
$x_{1} = x_{2} = y_{1} = y_{2} = z_{1} = z_{2} \equiv 1$. 
In the following, let $\mathbf{r}_i=( r_{ix},r_{iy})$ be the coordinate vector $(N_i,M_i)$ of associated 
to the unit cell containing site $i$, as defined in the main text. 
$G_{ij}$ is easily calculated from the Fourier transform \eqref{kas_matrix} of the matrix $A$, 
\begin{eqnarray}\label{g1}
G_{ij} 
&=& \int_{0}^{2\pi} \int_{0}^{2\pi} \frac{d\theta d\phi}{(2\pi)^{2}} 
e^{i[(r_{ix}-r_{jx})\theta+(r_{iy}-r_{jy})\phi]}
A^{-1}(\theta, \phi)_{ij}\nonumber\\
&=& \int_{0}^{2\pi} \int_{0}^{2\pi} \frac{d\theta d\phi}{(2\pi)^{2}} 
e^{i[\Delta N\cdot\theta+\Delta M\cdot\phi]}
A^{-1}(\theta, \phi)_{ij}\;,
\end{eqnarray}
where the inverse matrix $A^{-1}(\theta, \phi)$ of \eqref{kas_matrix} is given by 
\begin{eqnarray}\label{inv_x}
A^{-1}(\theta,\phi)  =  \frac{1}{4}   
\begin{pmatrix}
0                           &  a_{1}          & a_{2}     & a_{3} \cr
-a_{1}^{\star}              &  0              & a_{4}     & a_{5} \cr
-a_{2}^{\star}              &  -a_{4}^{\star} & 0         & a_{6} \cr
-a_{3}^{\star}              &  -a_{5}^{\star} & -a_{6}^{\star}     & 0\cr
\end{pmatrix}
\end{eqnarray}
with
\begin{eqnarray}\label{inv_x1}
	&& a_{1} = -(1 - e^{-i\theta})        \;\;\;\;\;\;\;\;\;\;\;\;\;\;  a_{4} = -(e^{i\theta} - e^{-i\phi}) \nonumber\\
	&& a_{2} = 1 + e^{-i\phi}             \,\;\;\;\;\;\;\;\;\;\;\;\;\;\;\;\;\;\;\;  a_{5} = -(1 + e^{-i\phi})        \\
	&& a_{3} = e^{-i\theta} - e^{-i\phi}  \;\;\;\;\;\;\;\;\;\;\;\;\;\;\;  a_{6} = -(1 - e^{-i\theta})\;.    \nonumber
\end{eqnarray}
\begin{figure}[t]
\includegraphics[width=0.99\textwidth]{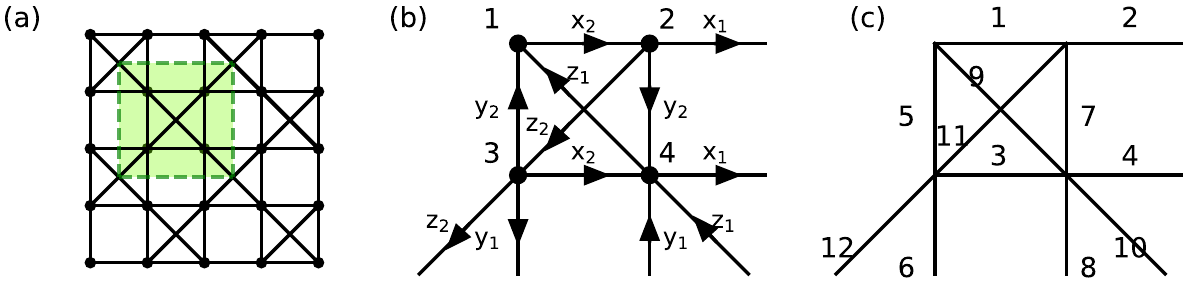}
\caption{
        (a) The checkerboard lattice is shown, the unit cell is depicted 
        inside the dashed lines. (b) shows the four-site unit cell of the 
        checkerboard lattice. It contains twelve links each equipped with a respective 
        weight of $x_{1}, x_{2}, y_{1}, y_{2}, z_{1}, z_{2}$. 
        The arrows on the links indicate the Kasteleyn orientation.  
        (c) The twelve links in the unit cell for that the dimer-dimer correlations are computed are labeled 
        $i = 1, \ldots, 12$. 
}
\label{figure1_666abc}
\end{figure}	

Moreover, plugging \eqref{inv_x} into \eqref{g1} results into an important observation. 
The matrix elements in \Eq{inv_x1} have $e^{i\theta}$ and $e^{i\phi}$ show up only with 
limited powers, in this case $-1$, $0$, and $1$. This is a rather special feature of the present lattice geometry. 
It leads to the fact  that $G_{ij}$ vanishes for $|r_{ix} - r_{jx}| \geq 2$ and for 
$|r_{1y} - r_{2y}| \geq 2$. 
Consequently, the connected correlation function 
$C[ij,kl]$ 
{\it vanishes identically} if the distance between the two lattice links under consideration is 
larger than two lattice spacings, 
\begin{eqnarray}\label{corrvan}
	C[ij,kl] = 0 \;\;\;\text{if}\;\;\; \mbox{$|r_{ix} - r_{jx}| \geq 2$ or 
$|r_{1y} - r_{2y}| \geq 2$.}
\end{eqnarray}	
This vanishing of correlations beyond a certain distance is a rather special property only shared by a few 
(quantum) dimer models on the kagome \cite{misguich, wang2007exact,wu2008dimers, Oshikawa}
and related  \cite{loh2008dimers} geometries. It is indicative of a deeper solvable structure 
of the present model that these other cases also have, as outlined in the discussion section of the main text. 
The present findings show that this underlying solvable structure may also exist in the non-planar case. 

We may compute expectation values of the dimer link variable $n_{ij}$ and all its nonvanishing correlation functions 
by utilizing \eqref{g1}, \eqref{inv_x} and \eqref{corrs2a}. 
This includes correlations between two links within the same unit cell as well as two links in two neighboring unit cells. 
For the single-link expectation values, we obtain 
\begin{eqnarray}\label{single_n}
\langle n_{i} \rangle = \left\{\begin{array}{ll}
                1/8 & \text{for all horizontal and vertical links }i = 1,\ldots,8\\
                1/4 & \text{for all diagonal links  }i = 9,\ldots,12
\end{array}\right.
\end{eqnarray}
where the numbers refer to \Fig{figure1_666}$(c)$. 
As expected, this reflects all the symmetries of the lattice, which in particular includes $\pi/2$-rotation about each plaquette-center. 
Next, we compute the connected correlation function for the three cases that (I) both dimers $i$ and $j$ live 
on links in the same unit cell $\mathbf{r}= (0,0)$, (II) one dimer is located at position 
$\mathbf{r}_{1} = (0,0)$ while the second dimer is located in the horizontally adjacent unit cell 
$\mathbf{r}_{2} = (1,0)$, and (III) one dimer is placed at position 
$\mathbf{r}_{1} = (0,0)$ while the second dimer is located at $\mathbf{r}_{2} = (0,1)$. 
The results are given in 
Tables ~\ref{table1}-~\ref{table3}. 
 
\begin{table}[h!]
\begin{tabular}{r|rrrrrrrrrrrr}
\hline\hline
\multicolumn{13}{c}{$(r_y,r_x)=(\Delta M, \Delta N)=(0,0)$}\\ \hline
    &  1    &  2    &  3    &  4    &  5    &  6    &  7    &  8    &  9    &  10   &  11   &  12   \\ \hline
 1  &  $7a$ &  $-a$ &  $-a$ &  $a$  &  $-a$ &  $a$  &  $-a$ &  $a$  &  $-b$ &  $b$  &  $-b$ &  $b$  \\
 2  &  $-a$ &  $7a$ &  $a$  &  0    &  $b$  &  0    &  $-a$ &  0    &  0    &  0    &  $-b$ &  0    \\
 3  &  $-a$ &  $a$  &  $7a$ &  $-a$ &  $-a$ &  $-a$ &  $-a$ &  $-a$ &  $-b$ &  $-b$ &  $-b$ &  $-b$ \\
 4  &  $a$  &  0    &  $-a$ &  $7a$ &  $a$  &  0    &  $-a$ &  $-a$ &  $-b$ &  $-b$ &  0    &  0    \\
 5  &  $-a$ &  $b$  &  $-a$ &  $a$  &  $7a$ &  $-a$ &  $-a$ &  $a$  &  $-b$ &  $b$  &  $-b$ &  $-b$ \\
 6  &  $a$  &  0    &  $-a$ &  0    &  $-a$ &  $7a$ &  $a$  &  0    &  0    &  0    &  $-b$ &  $-b$ \\
 7  &  $-a$ &  $-a$ &  $-a$ &  $-a$ &  $-a$ &  $a$  &  $7a$ &  $-a$ &  $-b$ &  $-b$ &  $-b$ &  $b$  \\
 8  &  $a$  &  0    &  $-a$ &  $-a$ &  $a$  &  0    &  $-a$ &  $7a$ &  $-b$ &  $-b$ &  0    &  0    \\
 9  &  $-b$ &  0    &  $-b$ &  $-b$ &  $-b$ &  0    &  $-b$ &  $-b$ &  $3c$ &  $-c$ &  $c$  &  0    \\
10  &  $b$  &  0    &  $-b$ &  $-b$ &  $b$  &  0    &  $-b$ &  $-b$ &  $-c$ &  $3c$ &  0    &  0    \\
11  &  $-b$ &  $-b$ &  $-b$ &  0    &  $-b$ &  $-b$ &  $-b$ &  0    &  $c$  &  0    &  $3c$ &  $-c$ \\
12  &  $b$  &  0    &  $-b$ &  0    &  $-b$ &  $-b$ &  $b$  &  0    &  0    &  0    &  $-c$ &  $3c$ \\
\hline\hline
\end{tabular}
\caption{Correlation function between two dimers in the 
same unit cell. We define $a = 1/64$, $ b = 1/32$ and $c = 1/16$ (see \Fig{figure1_666abc}$(c)$). 
}\label{table1}
\end{table}
\begin{table}
\begin{tabular}{r|rrrrrrrrrrrr}
\hline\hline
\multicolumn{13}{c}{$(r_y,r_x)=(\Delta M, \Delta N)=(0,1)$}\\ \hline
    &  1   &  2   &  3                     &  4   &  5   &  6   &  7                    &  8   &  9   & 10                     & 11   & 12   \\ \hline
 1  &  0   &  0   &  {\color{white}{-}}0   &  0   &  0   & $-a$ &  {\color{white}{-}}0  &  0   &  0   &  {\color{white}{-}}0   &  0   &  0   \\
 2  &  0   &  0   &  {\color{white}{-}}$a$ &  0   & $-a$ &  0   &  $a$                  &  0   & $-b$ &  0                     &  0   &  0   \\
 3  &  0   &  $a$ &  0                     &  0   &  0   &  $a$ &  0                    &  0   &  0   &  0                     &  0   &  0   \\
 4  &  0   &  0   &  0                     &  0   & $-a$ & $-a$ &  $a$                  &  0   &  0   &  0                     & $-b$ & $-b$ \\
 5  &  0   & $-a$ &  0                     & $-a$ &  0   & $-a$ &  0                    &  0   &  0   &  0                     &  0   &  0   \\
 6  & $-a$ &  0   &  $a$                   & $-a$ & $-a$ &  0   &  0                    &  0   &  0   &  0                     &  0   &  0   \\
 7  &  0   &  $a$ &  0                     &  $a$ &  0   &  0   &  0                    &  0   &  0   &  0                     &  0   &  0   \\
 8  &  0   &  0   &  0                     &  0   &  0   &  0   &  0                    &  0   &  0   &  0                     &  $b$ & $-b$ \\
 9  &  0   & $-b$ &  0                     &  0   &  0   &  0   &  0                    &  0   &  0   &  0                     &  0   &  0   \\
10  &  0   &  0   &  0                     &  0   &  0   &  0   &  0                    &  0   &  0   &  0                     &  0   &  $c$ \\
11  &  0   &  0   &  0                     & $-b$ &  0   &  0   &  0                    &  $b$ &  0   &  0                     &  0   &  0   \\
12  &  0   &  0   &  0                     & $-b$ &  0   &  0   &  0                    & $-b$ &  0   &  $c$                   &  0   &  0   \\
\hline\hline
\end{tabular}
\caption{Correlation function between two dimers in horizontally adjacent unit cells 
$(\mathbf{r}_1 = (0,0)$ and $\mathbf{r}= (1,0)$. 
Row indices $1,2,\dots, 12$ label links in unit cell $(0,0)$ while 
column indices label links in unit cell $(1,0)$. 
We again define $a = 1/64$, $ b = 1/32$ and $c = 1/16$ (see \Fig{figure1_666abc}$(c)$).  
} \label{table2}
\end{table}
\begin{table}
\begin{tabular}{r|rrrrrrrrrrrr}
\hline\hline
\multicolumn{13}{c}{$(r_y,r_x)=(\Delta M, \Delta N)=(1,0)$}\\ \hline
    &  1    &  2    &  3                   &  4    &  5    &  6    &  7                     &  8    &  9    & 10                   & 11    & 12                   \\ \hline
 1  &  0    &  $-a$ &  {\color{white}{-}}0 &  0    &  0    &  0    &  0                     &  0    &  0    &  {\color{white}{-}}0 &  0    &  {\color{white}{-}}0 \\
 2  &  $-a$ &  0    &  0                   &  0    &  0    &  0    &  0                     &  0    &  0    &  0                   &  0    &  0                   \\
 3  &  0    &  0    &  0                   &  0    &  0    &  0    &  0                     &  0    &  0    &  0                   &  0    &  0                   \\
 4  &  0    &  0    &  0                   &  0    &  $-a$ &  0    &  {\color{white}{-}}$a$ &  0    &  0    &  0                   &  $b$  &  0                   \\
 5  &  0    &  0    &  0                   &  $-a$ &  0    &  0    &  0                     &  0    &  0    &  0                   &  0    &  0                   \\
 6  &  0    &  0    &  0                   &  0    &  0    &  0    &  $a$                   &  0    &  $-b$ &  0                   &  0    &  0                   \\
 7  &  0    &  0    &  0                   &  $a$  &  0    &  $a$  &  0                     &  0    &  0    &  0                   &  0    &  0                   \\
 8  &  0    &  0    &  0                   &  0    &  0    &  0    &  0                     &  0    &  0    &  0                   &  $-b$ &  0                   \\
 9  &  0    &  0    &  0                   &  0    &  0    &  $-b$ &  0                     &  0    &  0    &  0                   &  0    &  0                   \\
10  &  0    &  0    &  0                   &  0    &  0    &  0    &  0                     &  0    &  0    &  0                   &  0    &  0                   \\
11  &  0    &  0    &  0                   &  $b$  &  0    &  0    &  0                     &  $-b$ &  0    &  0                   &  0    &  0                   \\
12  &  0    &  0    &  0                   &  0    &  0    &  0    &  0                     &  0    &  0    &  0                   &  0    &  0                   \\
\hline\hline
\end{tabular}
\caption{Correlation function between two dimers in vertically adjacent unit cells 
$\mathbf{r}_1 = (0,0)$ and $\mathbf{r}_2= (0,1)$. 
Again row indices $1,2,\dots, 12$ label links in unit cell $(0,0)$ while 
column indices label edges in unit cell $(0,1)$. 
We define $a = 1/64$ and $ b = 1/32$ (see \Fig{figure1_666abc}$(c)$). 
} \label{table3}
\end{table}
\end{widetext}
\end{document}